\documentclass[sigconf,nonacm]{acmart}
\usepackage{tabularx}
\AtBeginDocument{%
  \providecommand\BibTeX{{%
    \normalfont B\kern-0.5em{\scshape i\kern-0.25em b}\kern-0.8em\TeX}}}
\usepackage[export]{adjustbox}
\makeatletter
\def\@copyrightspace{\relax}
\makeatother
\begin{document}
\title{Streamlining CXL Adoption for Hyperscale Efficiency}

\author{Angelos Arelakis}
\authornotemark[1]
\author{Nilesh Shah}
\authornotemark[1]
\author{Yiannis Nikolakopoulos}
\authornotemark[1]
\author{Dimitrios Palyvos-Giannas}
\email{<first_name>@zptcorp.com}
\authornote{All authors contributed equally to this research.}

\affiliation{%
  \institution{ZeroPoint Technologies AB,~~~Gothenburg}
  \country{Sweden}
}
\begin{abstract}
In our exploration of Composable Memory systems utilizing CXL, we focus on overcoming adoption barriers at Hyperscale, underscored by economic models demonstrating Total Cost of Ownership (TCO). While CXL addresses the pressing memory capacity needs of emerging Hyperscale applications, the escalating demands from evolving use cases such as AI outpace the capabilities of current CXL solutions. Hyperscalers resort to software-based memory (de)compression technology, alleviating memory capacity, storage, and network constraints but incurring a notable "Tax" on Compute CPU cycles.
As a pivotal guide to the CXL community, Hyperscalers have formulated the groundbreaking Open Compute Project (OCP) Hyperscale CXL Tiered Memory Expander specification. If implemented, this specification lowers TCO adoption barriers, enabling diverse CXL deployments at both Hyperscaler and Enterprise levels. We present a CXL integrated solution, aligning with the aforementioned specification, introducing an energy-efficient, scalable, hardware-accelerated, Lossless Compressed Memory CXL Tier. This solution, slated for mid-2024 production and open for integration with Memory Expander controller manufacturers, offers 2-3X CXL memory compression in nanoseconds, delivering a 20-25\%
reduction in TCO for end customers without requiring additional physical slots.
In our discussion, we pinpoint areas for collaborative innovation within the CXL Community to expedite software/hardware advancements for CXL Tiered Memory Expansion. Furthermore, we delve into unresolved challenges in Pooled deployment and explore potential solutions, collectively aiming to make CXL adoption a "No Brainer" at Hyperscale.
\end{abstract}





\maketitle

\section{Introduction }
\subsection{Background}
DRAM is a key driver of performance and cost in the Data Center. CXL offers a path towards improved DRAM utilization (cost efficiency) in both Tiered and Pooled scenarios, but the Total Cost of system Ownership (TCO) increases non trivially due to added infrastructure components~\cite{10.1109/MM.2023.3241586}. Moving/adding DRAM off the DDR interface to CXL requires significant return on investment and increased efficiency. The ground breaking OCP specification for the Hyperscale CXL Tiered Memory Expander offers a crucial solution to achieve efficiency by specifying the addition of a compressed DRAM tier. Compression increases the effective capacity of a CXL memory device, reducing the power necessary to manufacture and operate DRAM components. 
\subsection{OCP Specification/ requirements}
The OCP Specification~\cite{OCPSpecification} calls for a sustainable, \textit{transparent} and cost-efficient method to compress memory on CXL Type 3 devices on a variety of compute platforms with a diversity of memory technologies. The OCP Spec calls for Access to a cache line in a compressed block within 250ns, with tail latency for accessing a cache line in an compressed block of <1us including worst case lookup latency, decompression, power-state transitions. Furthermore, decompression speeds of 46GB/s must be matched to 4 Channels at 1867MT/s compressed data, with 4kB/1kB blocks. State of the Art solutions fail to meet these requirements.
\subsection{State of the Art}
General-purpose, software-based lossless data “(de)compression” techniques are used widely in hyperscale systems to alleviate the memory, storage, and network cost with significant associated compute overheads “datacenter taxes" in warehouse-scale datacenter services~\cite{10.1145/3579371.3589074,10158161}. These (de)compression services consume 2.9\% to 4.6\% of fleet CPU cycles and 10-50\% cycles in key services. The existing (de)compression algorithm implementations (at block size granularity) do not meet the latency, power efficiency and preservation of software investment tenets spelled out in the OCP Specification. Waiting for an entire block or page prior to decompress data incurs latency measured in microseconds, which is suitable for storage but not acceptable for memory performance. Software and Intel QAT (De)compression solutions require the use of area and power intensive Xeon cores~\cite{IntelQAT}, falling order of  magnitude short of the latency and bandwidth requirements specified in the OCP Spec. State of the Art hardware accelerators reported in open source (Zipline) or in other studies (CDPU) reported consuming 1.3-5.7mm sq area while achieving 5-11GB/s but with unspecified latency and performance scalability~\cite{10.1145/3579371.3589074}. These designs are not geared /portable for integration into CXL Tiered Memory Expander devices as a complete solution to the best of our knowledge.
\section{Solution}
\subsection{Invention}
\begin{figure}
     \centering
     \includegraphics[width=.9\linewidth]{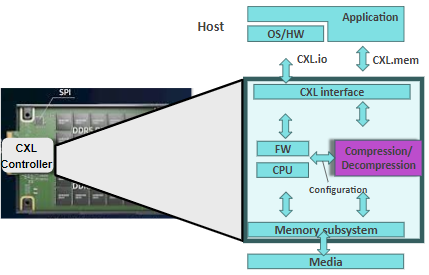}
     \caption{(De)Compression Solution integrated into CXL Expander Type 3 Device}
     \label{fig:soln}
 \end{figure}

We present our Hardware accelerated, Lossless memory (De) Compression integrated IP solution [Figure~\ref{fig:soln}] that increases the effective CXL Type 3 Device memory capacity by 2x through transparent, in-line memory compression/decompression at 64 byte cache line granularity. It is designed as an area and power-efficient solution composed of multiple integrated IP blocks, portable across the latest process nodes, supporting (LP)DDR4 and (LP)DDR5 memory technologies. This solution meets the latency and bandwidth requirements of the OCP Hyperscale CXL Tiered Memory expander spec by implementing a proprietary (De)compression algorithm at cache line granularity, with dual hardware accelerator implementation of the open source LZ4 algorithm to operate at page or block granularity for legacy compatibility.

\begin{figure}
    \centering
    \includegraphics[width=1\linewidth]{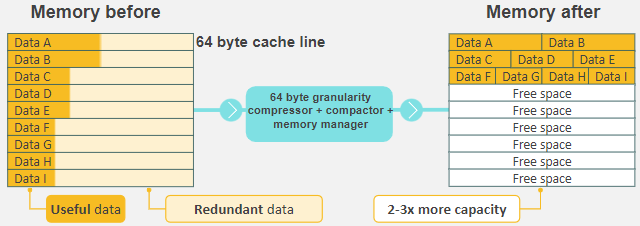}
    \caption{Cache line (64 byte) compression algorithm}
    \label{fig:64b}
\end{figure}
\vspace{-1.8mm}

\subsection{Innovation}
Unlike state-of-the-art solutions that conduct compression at block/ page granularity, our approach operates at a more refined 64-byte granularity (as illustrated in Figure \ref{fig:64b}). This innovation seamlessly integrates into the CXL Type 3 device System-on-Chip (SoC), supporting both AXI4 and CHI specifications. The communication with the device occurs over CXL (2.0, 3.0, 3.1) command sets translated to AXI4, operating at a frequency of 1.2GHz (@4nm Samsung). Our solution exposes a compressed memory region as an additional NUMA tier in the memory hierarchy.

Achieving a 2-3x effective capacity increase with minimal impact on device bandwidth and nanosecond-level latency, our solution dynamically manages the compressed memory tier within the CXL Type 3 device. It implements real-time compression/decompression with compaction, operating at main memory speed and throughput. Additionally, an optional adaptive feature tunes performance based on diverse workloads using low-level telemetry data.

When the host processor demotes pages from the directly-connec-ted DRAM to the CXL device, the CXL (device) controller converts CXL.mem data to AXI/CHI commands and calls our IP block in real time. The IP block intercepts data at cache line granularity (with a proprietary (de)compression algorithm) or at page/block granularity, as instructed over CXL.io. The compression algorithm is then applied, and memory allocation for the compressed CXL tier is managed. Compressed data is stored in the DRAM memory media over the DDR interface, with the IP block situated between the CXL controller interface and the DDR controller interface, communicating via AXI4 or CHI.

Our solution not only performs compression but also conducts compaction, ensuring effective space utilization. The IP block implements capacity reporting telemetry to the host. When a compressed page is requested, the IP accesses the compressed cache lines or pages/blocks, decompresses the data, performs necessary address translation on-the-fly, and transmits the data over the AXI interconnect from DRAM media to the CXL interface. This entire process is orchestrated by lightweight firmware running on an embedded processor in the CXL Type 3 expander SoC, enabling real-time compression/decompression at CXL line speed.
\begin{figure}
    \centering
    \includegraphics[width=0.9\linewidth]{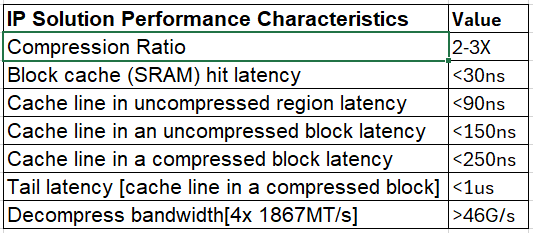}
    \vspace{-1mm}
    \caption{IP Solution Performance Characteristics. Exceed OCP Specification requirements}
    \label{fig:table}
\end{figure}

\subsection{Performance Summary}
\begin{figure}
    \centering
    \includegraphics[width=1\linewidth]{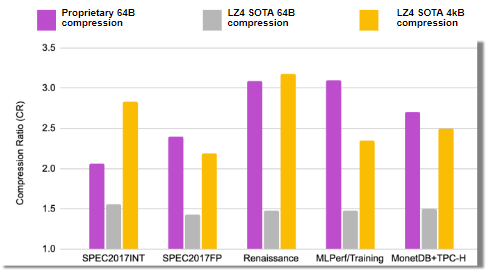}
    \caption{Geomean Compression Ratio 
across applications of each dataset}
    \label{fig:perf}
\end{figure}

\begin{figure}
    \centering
    \includegraphics[width=1\linewidth]{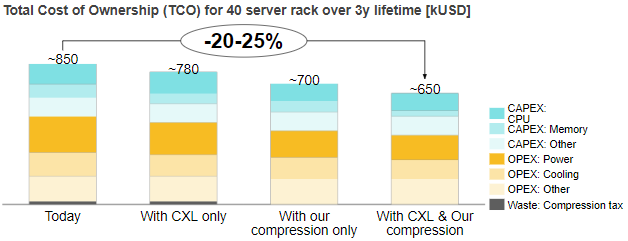}
    \caption{Reduced TCO. Addition of Compressed Memory Tier  increases effective capacity (GB) leading to reduced \$/GB  }
    \label{fig:tco}
\end{figure}

The IP Solution described above delivers\textbf{ 2-3X Compression Ratios} while compressing memory at \textbf{Cache Line (64 Byte) granularity} across a breadth of representative Data Center workloads including \textbf{SPEC2017INT/FP, Renaissance, MLPerf/ Training, MonetDB+TPC-H}  [Figure\ref{fig:perf}]. The solution has been verified to operate at 1.2GHz and fits in an area of approximately \textbf{0.9mm\(^2\) (at 4nm Samsung) where 75\% of the IP solution area is occupied by SRAM}. This solution currently \textbf{supports (LP)DDR4, (LP)DDR5} memory technologies and \textbf{decompresses data in single digit clock cycle latency}. The overall CXL IP solution performance characteristics [Figure \ref{fig:table}] satisfy the requirements in the OCP Spec \cite{OCPSpecification}. The key IP Solution performance characteristics are in Figure \ref{fig:table}. The IP solution is slated to go into production mid 2024 into CXL controller products and is expected to provide savings of 20-25\% TCO [Figure  \ref{fig:tco}] using previously published TCO model assumptions  \cite{ElasticsTCO}, factoring in 2X capacity expansion through an additional Compressed Memory Tier. 

\begin{figure}
    \centering
    \includegraphics[width=0.25\linewidth]{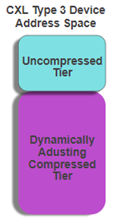}
    \caption{Compressed Memory Tier introduced and managed by the IP Solution within the CXL Memory Expander Type 3 device controller}
    \label{fig:compressed memory tier }
\end{figure}
\section{Implementation / Evaluation}
\subsection{Proof of Concept Demonstration}
\begin{figure}
    \centering
    \includegraphics[width=.85\linewidth]{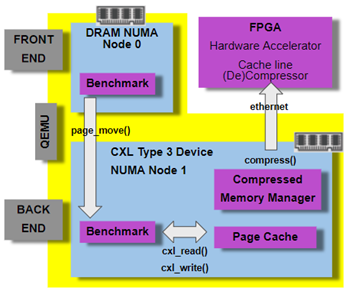}
    \caption{Proof Of Concept Implementation Demonstration }
    \label{fig:fpga}
\end{figure}

To validate the practical implementation and effectiveness of the proposed CXL solution, a Proof of Concept (PoC) demonstration has been developed [Figure \ref{fig:fpga}]. The demonstration comprises two integral components: the frontend, based on QEMU, emulating the host system and managing the migration of identified cold pages to the CXL subsystem, and the backend, featuring an inline memory compression/decompression and management accelerator implemented on an FPGA, along with its accompanying firmware running on the FPGA-based platform.
\vspace{-1mm}

\subsection{Frontend: QEMU-based Emulation}
The QEMU frontend emulates a Linux host capable of running realistic workloads. Currently, arbitrary pages of the workload are migrated into the emulated CXL memory. This can be extended with several mechanisms for detecting cold pages (e.g. LRU or MGLRU algorithms, Senpai, DAMON).
\subsection{Backend: FPGA-based Accelerator and Firmware}
The backend is centered around the FPGA, hosting an inline memory compression/decompression and management accelerator. This FPGA-based solution, coupled with its firmware, is responsible for compressing, storing, and managing the data migrated from the host system. The FPGA acts as the compressed memory subsystem of the CXL device, encapsulating both hardware and firmware components.
\subsection{Demonstration Workflow}
\begin{figure}
    \centering
    \includegraphics[width=1\linewidth]{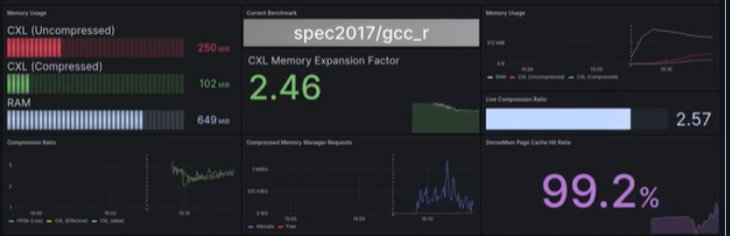}
    \caption{FPGA Proof of Concept. Dynamically adjusting CXL Compressed Memory Tier}
    \label{fig:telemetry}
\end{figure}

\begin{enumerate}
    \item \textbf{Migration of Pages}: The QEMU frontend migrates pages to the CXL subsystem. This process involves invoking the IP solution on the FPGA, which takes on the responsibilities of compression, storage, and bookkeeping of the data.
    \item \textbf{Requesting Compressed Pages}: The QEMU frontend, representing the host system, requests the migration of compressed pages by sending corresponding read requests to the IP solution on the FPGA. This step simulates the real-world scenario where the host system interacts with the CXL device to access compressed data.
    \item \textbf{Real Application Utilization}: The demonstration includes a real application running on the system, showcasing the practical advantages of an expanded memory on the FPGA. This expanded memory acts as the compressed memory subsystem of the CXL device, incorporating both hardware and firmware functionalities.
    \item \textbf{Live Telemetry Display}: Throughout the demonstration, live telemetry data is presented, including compression ratios, expansion factors, and other metrics [Figure \ref{fig:telemetry}] aligned with the OCP requirements for Hyperscale CXL Tiered Memory Expander. This real-time feedback provides insight into the system's performance and adherence to industry standards.
\end{enumerate}

\vspace{-1mm}

The PoC demonstration serves as a tangible representation of the proposed CXL solution's capabilities. By combining emulation with FPGA-based hardware acceleration, the demonstration provides a holistic view of the solution's functionality, validating its potential for real-world applications in memory-intensive scenarios. 

\subsection{Summary}
CXL addresses the pressing memory capacity needs of emerging Hyperscale applications. Hyperscalers~\cite{OCPSpecification} have formulated the groundbreaking (OCP) Hyperscale CXL Tiered Memory Expander specification to lower TCO and enable diverse CXL deployments. This paper presents a CXL integrated solution, aligning with
the aforementioned specification, introducing an energy-efficient, scalable, hardware-accelerated, Lossless Compressed Memory CXL Tier offering 2-3X CXL memory compression in nanoseconds and delivering a 20-25\% reduction in TCO for end customers without requiring additional physical slots. 

\section{Open Questions / Call for Community Collaboration}
\begin{enumerate}
    \item \textbf{Upstream Linux Driver Development}: Collaboration within the CXL community extends to upstream Linux driver development. A lightweight Linux driver is under development and is set to be upstreamed to the Linux kernel. This driver provides APIs that allow the host to utilize the oversubscribed compressed memory region of the device, and is intended to be fully compliant with the CXL upstream Linux driver. Collaboration in this area will ensure seamless integration with the Linux ecosystem, providing a standardized interface for CXL memory expansion.
    \item \textbf{Integration and Testing/Benchmarking}: To ensure the robustness and performance of the proposed CXL solution, collaboration efforts could be directed towards integration and testing/benchmarking with data center applications, operating systems, and hypervisors. A comprehensive testing framework will be established to validate the solution's compatibility and efficiency across diverse environments, addressing the unique requirements of different workloads.
    \item \textbf{Collaboration with Hyperscalers and Device Manufacturers}: Critical to the success of CXL adoption is collaboration with hyperscalers and device manufacturers. Aligning on requirements, gathering feedback, and understanding the practical challenges faced by these stakeholders are paramount to ensure smooth production deployment. Collaborative efforts involve iterative discussions, prototype testing, and refining solutions to ensure they meet the specific needs of hyperscalers and device manufacturers.
    \item \textbf{Addressing adoption challenges raised by Hyperscalers}: Hyperscaler End customers have recently argued against the value of implementing CXL memory pools~\cite{10.1145/3626111.3628195}, their main arguments being: The cost of a CXL pool will outweigh any savings from reducing RAM. CXL has substantially higher latency than main memory, enough so that using it will require substantial rewriting of network applications in complex ways. To reduce Total Cost of Ownership (TCO) and enhance CXL adoption, the community can convert this challenge into an opportunity to path find better together solutions with compression, optical interconnects, native CXL controllers, and integrating flash media into the CXL memory tier to create holistic solutions that address diverse aspects of CXL infrastructure to maximize efficiency and  performance at hyper scale. The open-source community in particular has the opportunity to take leadership, rally around to develop solutions that require no fundamental programming changes, ensuring a smooth transition for existing software stacks.
\end{enumerate}

\bibliographystyle{ACM-Reference-Format}
  \bibliography{sample-acmtog}


\begin{thebibliography}{7}


\ifx \showCODEN    \undefined \def \showCODEN     #1{\unskip}     \fi
\ifx \showDOI      \undefined \def \showDOI       #1{#1}\fi
\ifx \showISBNx    \undefined \def \showISBNx     #1{\unskip}     \fi
\ifx \showISBNxiii \undefined \def \showISBNxiii  #1{\unskip}     \fi
\ifx \showISSN     \undefined \def \showISSN      #1{\unskip}     \fi
\ifx \showLCCN     \undefined \def \showLCCN      #1{\unskip}     \fi
\ifx \shownote     \undefined \def \shownote      #1{#1}          \fi
\ifx \showarticletitle \undefined \def \showarticletitle #1{#1}   \fi
\ifx \showURL      \undefined \def \showURL       {\relax}        \fi
\providecommand\bibfield[2]{#2}
\providecommand\bibinfo[2]{#2}
\providecommand\natexlab[1]{#1}
\providecommand\showeprint[2][]{arXiv:#2}

\bibitem[Apostol(2022)]%
        {ElasticsTCO}
\bibfield{author}{\bibinfo{person}{George Apostol}.} \bibinfo{year}{2022}\natexlab{}.
\newblock \bibinfo{booktitle}{\emph{Using Pools of Shared Resources to Lower Latency and Improve System Performance}}.
\newblock
\urldef\tempurl%
\url{https://drive.google.com/file/d/1cZGC64WFY491-Jrf7jAHy64xR-YyDIOD/view}
\showURL{%
\tempurl}


\bibitem[Berger et~al\mbox{.}(2023)]%
        {10.1109/MM.2023.3241586}
\bibfield{author}{\bibinfo{person}{Daniel~S. Berger}, \bibinfo{person}{Daniel Ernst}, \bibinfo{person}{Huaicheng Li}, \bibinfo{person}{Pantea Zardoshti}, \bibinfo{person}{Monish Shah}, \bibinfo{person}{Samir Rajadnya}, \bibinfo{person}{Scott Lee}, \bibinfo{person}{Lisa Hsu}, \bibinfo{person}{Ishwar Agarwal}, \bibinfo{person}{Mark~D. Hill}, {and} \bibinfo{person}{Ricardo Bianchini}.} \bibinfo{year}{2023}\natexlab{}.
\newblock \showarticletitle{Design Tradeoffs in CXL-Based Memory Pools for Public Cloud Platforms}.
\newblock \bibinfo{journal}{\emph{IEEE Micro}} \bibinfo{volume}{43}, \bibinfo{number}{2} (\bibinfo{date}{mar} \bibinfo{year}{2023}), \bibinfo{pages}{30–38}.
\newblock
\showISSN{0272-1732}
\urldef\tempurl%
\url{https://doi.org/10.1109/MM.2023.3241586}
\showDOI{\tempurl}


\bibitem[Chauhan et~al\mbox{.}(2023)]%
        {OCPSpecification}
\bibfield{author}{\bibinfo{person}{Meta:~Prakash Chauhan}, \bibinfo{person}{Chris Petersen}, \bibinfo{person}{Google:~Brian Morris}, {and} \bibinfo{person}{Jerome Glisse}.} \bibinfo{year}{2023}\natexlab{}.
\newblock \bibinfo{booktitle}{\emph{OCP Hyperscale CXL Tiered Memory Expander Specification,Revision 1 Version 1.0 Base Specification, Template v1.2, Effective October 27, 2023}}.
\newblock
\urldef\tempurl%
\url{https://www.opencompute.org/documents/hyperscale-cxl-tiered-memory-expander-for-ocp-base-specification-1-pdf}
\showURL{%
\tempurl}


\bibitem[Jeong et~al\mbox{.}(2023)]%
        {10158161}
\bibfield{author}{\bibinfo{person}{Geonhwa Jeong}, \bibinfo{person}{Bikash Sharma}, \bibinfo{person}{Nick Terrell}, \bibinfo{person}{Abhishek Dhanotia}, \bibinfo{person}{Zhiwei Zhao}, \bibinfo{person}{Niket Agarwal}, \bibinfo{person}{Arun Kejariwal}, {and} \bibinfo{person}{Tushar Krishna}.} \bibinfo{year}{2023}\natexlab{}.
\newblock \showarticletitle{Characterization of Data Compression in Datacenters}. In \bibinfo{booktitle}{\emph{2023 IEEE International Symposium on Performance Analysis of Systems and Software (ISPASS)}}. \bibinfo{pages}{1--12}.
\newblock
\urldef\tempurl%
\url{https://doi.org/10.1109/ISPASS57527.2023.00010}
\showDOI{\tempurl}


\bibitem[Karandikar et~al\mbox{.}(2023)]%
        {10.1145/3579371.3589074}
\bibfield{author}{\bibinfo{person}{Sagar Karandikar}, \bibinfo{person}{Aniruddha~N. Udipi}, \bibinfo{person}{Junsun Choi}, \bibinfo{person}{Joonho Whangbo}, \bibinfo{person}{Jerry Zhao}, \bibinfo{person}{Svilen Kanev}, \bibinfo{person}{Edwin Lim}, \bibinfo{person}{Jyrki Alakuijala}, \bibinfo{person}{Vrishab Madduri}, \bibinfo{person}{Yakun~Sophia Shao}, \bibinfo{person}{Borivoje Nikolic}, \bibinfo{person}{Krste Asanovic}, {and} \bibinfo{person}{Parthasarathy Ranganathan}.} \bibinfo{year}{2023}\natexlab{}.
\newblock \showarticletitle{CDPU: Co-designing Compression and Decompression Processing Units for Hyperscale Systems}. In \bibinfo{booktitle}{\emph{Proceedings of the 50th Annual International Symposium on Computer Architecture}} (Orlando, FL, USA) \emph{(\bibinfo{series}{ISCA '23})}. \bibinfo{publisher}{Association for Computing Machinery}, \bibinfo{address}{New York, NY, USA}, Article \bibinfo{articleno}{39}, \bibinfo{numpages}{17}~pages.
\newblock
\showISBNx{9798400700958}
\urldef\tempurl%
\url{https://doi.org/10.1145/3579371.3589074}
\showDOI{\tempurl}


\bibitem[Levis et~al\mbox{.}(2023)]%
        {10.1145/3626111.3628195}
\bibfield{author}{\bibinfo{person}{Philip Levis}, \bibinfo{person}{Kun Lin}, {and} \bibinfo{person}{Amy Tai}.} \bibinfo{year}{2023}\natexlab{}.
\newblock \showarticletitle{A Case Against CXL Memory Pooling}. In \bibinfo{booktitle}{\emph{Proceedings of the 22nd ACM Workshop on Hot Topics in Networks}} \emph{(\bibinfo{series}{HotNets '23})}. \bibinfo{publisher}{Association for Computing Machinery}, \bibinfo{address}{New York, NY, USA}, \bibinfo{pages}{18–24}.
\newblock
\showISBNx{9798400704154}
\urldef\tempurl%
\url{https://doi.org/10.1145/3626111.3628195}
\showDOI{\tempurl}


\bibitem[Will(2023)]%
        {IntelQAT}
\bibfield{author}{\bibinfo{person}{Brian Will}.} \bibinfo{year}{2023}\natexlab{}.
\newblock \bibinfo{booktitle}{\emph{Intel® QuickAssist Technology Zstandard Plugin, an External Sequence Producer for Zstandard}}.
\newblock
\urldef\tempurl%
\url{https://community.intel.com/t5/Blogs/Tech-Innovation/Artificial-Intelligence-AI/Intel-QuickAssist-Technology-Zstandard-Plugin-an-External/post/1509818}
\showURL{%
\tempurl}


\end{thebibliography}

\end{document}